\documentclass[paper]{JHEP3}
\usepackage{amssymb,enumerate}
\newcommand{\tmtexttt}[1]{{\ttfamily{#1}}}
\newenvironment{enumeratenumeric}{\begin{enumerate}[1.] }{\end{enumerate}}

\title{The POWHEG-hvq manual version 1.0}
\vfill
\author{Stefano Frixione \\
INFN, Sezione di Genova, Italy\\
E-mail: \email{Stefano.Frixione@cern.ch}}
\author{Paolo Nason \\
INFN, Sezione di Milano Bicocca, Italy\\
E-mail: \email{Paolo.Nason@mib.infn.it}}
\author{Giovanni Ridolfi \\
Dipartimento di Fisica, Universit\`a di Genova\\
and INFN, Sezione di Genova, Italy\\
E-mail: \email{Giovanni.Ridolfi@ge.infn.it}}
\abstract{
This note documents the use of the POWHEG-hvq package, a generator for
heavy flavour hadroproduction at next-to-leading order in QCD, that can be
easily interfaced to shower Monte Carlo programs, in such a way that NLO
and shower accuracy are both maintained.
}
\preprint{Bicocca-FT-07-10 \\GEF-TH-16/2007\\
}

\begin{document}

\section{Introduction}

The POWHEG-hvq program is a hard event generator for heavy quark production
in hadronic collisions. It is accurate at the next-to-leading order in QCD
(NLO from now on), and it can be interfaced to Shower Monte Carlo programs
(SMC from now on) like HERWIG and PYTHIA,
in such a way that both the leading logarithmic accuracy of the shower
and the NLO accuracy are maintained in the output. It is thus an
alternative to the MC@NLO heavy flavour production program
of ref.~\cite{Frixione:2003ei}.
The code can be found in
\begin{center}
 \tmtexttt{http://moby.mib.infn.it/\~{}nason/POWHEG}.
\end{center}
The program is an implementation of the heavy flavour NLO
cross sections of refs.~\cite{Nason:1988xz,Nason:1989zy,Mangano:1992jk},
according to the formalism of
refs.~\cite{Nason:2004rx} and \cite{Nason:2006hf}.
A detailed description of the implementation
is given in ref.~\cite{fnr2006}.
In the case of $t \bar{t}$ production, spin correlations in top decays are included
with a method similar to the one discussed in ref.~\cite{Frixione:2007zp}.
The relevant matrix elements for the spin correlations are the same
used in the MC@NLO
package \cite{Frixione:2006gn}, and were obtained using MadEvent \cite{Madevent}.
Our decay package takes into account the
finite width of the top quarks and of the $W$.

In this note, we give all the necessary information to run the
program.

\section{Installation}

The program comes as a tarred-gzipped file \tmtexttt{POWHEG-hvq.tar.gz}.
It can be installed with the following commands\\
\\
\tmtexttt{\$ tar -zxf POWHEG-hvq.tar.gz\\
\$ cd POWHEG-hvq\\
\$ make <target>}\\
\\
where the choice of the target depends upon the way one wants to interface the
program with a Shower Monte Carlo implementation.
The \tmtexttt{Makefile}
is set up to use the compiler \tmtexttt{g77} on Linux platforms.
If one wishes to use \tmtexttt{gfortran}, one
should uncomment the appropriate lines in the \tmtexttt{Makefile}.
Notice that some
versions of \tmtexttt{gfortran} may not support the \tmtexttt{idate} and
\tmtexttt{time} intrinsics. These are used in the \tmtexttt{mbook.f} file, \
that in turn is used in the examples. Thus, one may also need to comment out
the calls to \tmtexttt{idate} and \tmtexttt{time} in order to run the
examples with \tmtexttt{gfortran}.

\section{Modes of operation}
The program POWHEG-hvq generates hard events. The hard events can
then be fed into an SMC program 
for subsequent showering. POWHEG-hvq saves the hard event information
according to the Les Houches Interface for User Processes (LHIUP from now on)
conventions {\cite{Boos:2001cv}}. The SMC should also
comply with these conventions (as is the case for PYTHIA and HERWIG) in
order to be used in conjunction with POWHEG-hvq.

The program can be run in three ways.
\begin{itemize}
\item
POWHEG-hvq generates hard events, and store them in a file.
An SMC reads the file and showers it.
\item
POWHEG-hvq is linked directly to the SMC.
In this case the events are generated and immediately showered,
without intermediate storage.
\item
POWHEG-hvq is run as a standalone program, and the produced hard
events are analyzed without showering. The output yields in this case
NLO distribution with LL resummation of soft gluon effects.
\end{itemize}

\subsection{Storing the user events}\label{sec:lhef}
The easiest way to interface POWHEG-hvq to an SMC is to simply store the hard
events in a file (which we call the {\em event file}), 
and in a subsequent run read the events and process them with
the SMC. The format of the event file supported by POWHEG-hvq is the ``Standard
format for Les Houches event files'', documented in ref.~{\cite{Alwall:2006yp}}.
The program for the generation of the Les Houches
Event Files (LHEF from now on) can be built with the command\\
\\
\tmtexttt{\$ make main-lhef}\\
\\
The event file is named \tmtexttt{pwglhef.events} (the user is given the
possibility to change the file name, as documented in the next section). An
example program that reads the event file, showers it with HERWIG and
analyzes it the can be built as follows\\
\\
\tmtexttt{\$ make main-HERWIG-lhef}.\\
\\
A similar program, named
{\tt main-PYTHIA-lhef} is provided for {\tt PYTHIA}.
The HERWIG or PYTHIA program should be installed in the POWHEG-hvq directory.
In the case of HERWIG, the appropriate include files should also be present.
As can be evinced from the Makefile, the fortran
files relevant for these examples
are \tmtexttt{main-HERWIG-lhef.f}, \tmtexttt{herwig6510.f}
(\tmtexttt{main-PYTHIA-lhef.f}, \tmtexttt{pythia6326.f} for PYTHIA), 
\tmtexttt{mbook.f} and \tmtexttt{analize-hvq.f}.
The file \tmtexttt{analize-hvq.f} is a minimal analysis program,
provided as a starting example for more complex analysis.
It is adequate for
both HERWIG and PYTHIA (since it uses the standard common
blocks of ref.~\cite{Altarelli:1989wu}). It
uses the histogramming package of M.L.~Mangano, \tmtexttt{mbook.f}, and it
produces topdrawer output in the file {\tt pwgoutput.top}.
It can be used for charmed and bottomed hadron production, and for $t \bar{t}$
production. For charm and bottom, it analyses a few
kinematic observables, looking only at $D^{\pm}, D^0$,
$\overline{D^{}}^0$ for charm, and $B^{\pm}, B^0$, $\overline{B^{}}^0$
for bottom. For top pair production, the program looks for lepton pairs
with transverse momenta above a given cut.

POWHEG-hvq also provides an output that is as close as possible to the
MC@NLO user file format, so that users that have already implemented MC@NLO
in their analysis frameworks should be able to run POWHEG-hvq interfaced to
HERWIG with no extra work. The corresponding Makefile target is \
\tmtexttt{main-mcatnlofl}. The event file name is
\tmtexttt{pwgmcatnlofmt.events}.

\subsection{Interfacing POWHEG-hvq with a shower Monte Carlo program}

One should create a main program
that initializes the
SMC to make it ready to accept a user process,
and provide the following routines\\
\\
\tmtexttt{ \ \ \ \ \ subroutine UPINIT\\
\ \ \ \ \ call pwhginit\\
\ \ \ \ \ end\\
\\
\ \ \ \ \ subroutine UPEVNT\\
\ \ \ \ \ call pwhgevnt\\
\ \ \ \ \ end}\\
\\
that are the only link to the POWHEG-hvq program. The main program should
call the appropriate subroutines to run the SMC. If the SMC is compliant with
the LHIUP, it will call the routines  \tmtexttt{UPINIT} and \tmtexttt{UPEVNT}
in order to initialize, and to generate the hard events.
The routine {\tmtexttt pwhinit} performs the initialization of POWHEG-hvq,
setting up all the grids that are necessary for the efficient generation
of the events, and it also initializes the process common block of the LHIUP.
Each call to {\tmtexttt pwhgevnt} results in 
the generation of one event, and its storage
in the LHIUP event common block.

When using HERWIG, one must remove the dummy subroutines \tmtexttt{UPINIT} and
\tmtexttt{UPEVNT} that are present in the HERWIG source file.

We provide two examples, that can be built with the commands\\
\\
\tmtexttt{\$ make main-HERWIG}\\
\\
and\\
\\
\tmtexttt{\$ make main-PYTHIA}\\
\\
The analysis program is the same one as that
described in section~\ref{sec:lhef}.
\subsection{POWHEG-hvq as a standalone program}
In this case, the main program should have the structure\\
\\
\tmtexttt{ \ \ \ \ \ program MAIN\\
\ \ \ \ \ call pwhginit\\
\ \ \ \ \ do j=1,NEVENTS\\
\ \ \ \ \ \phantom{aaa} call pwhgevnt\\
c call some analysis routines here\\
\ \ \ \ \ \phantom{aaa}  ...\\
\ \ \ \ \ enddo\\
c Print out results\\
\ \ \ \ \ ...\\
\ \ \ \ \ end}\\
\\
No examples are provided. The analysis routines should make use of the
information stored in the LHIUP, as documented in ref.~\cite{Boos:2001cv}.

\section{Input parameters}

POWHEG-hvq provides an independent facility to set the input parameters for
the run. All parameters are stored in a file, named \tmtexttt{powheg.input}.
The format of the file is as follows
\begin{enumeratenumeric}
  \item Lines are no more than 100 characters long.
  
  \item Empty (blank) lines are ignored
  
  \item If a \tmtexttt{\#} or a \tmtexttt{!} appears at any point in a line,
  the part of the line starting from the \tmtexttt{\#} or \tmtexttt{!} symbol
  up to its end is blanked.
  
  \item An entry has the format:\\
  \tmtexttt{name \ \ value}\\
  usually followed by a \tmtexttt{!} and a comment to clarify the meaning of
  the variable. The \tmtexttt{name} keyword has no more than 20
  characters, and \tmtexttt{value} is an integer or floating point number.
  
  \item A maximum of 100 keywords are allowed.
\end{enumeratenumeric}
If the file \tmtexttt{powheg.input} is not present, the program asks the user
to enter a prefix, and then looks for the file
\tmtexttt{<prefix>-powheg.input}. In this case,
all the files created by POWHEG-hvq in
the current run will carry the prefix \tmtexttt{<prefix>-} instead of
\tmtexttt{pwg}.

The input parameters are read by the (\tmtexttt{real * 8}) function
\tmtexttt{powheginput(string)}, in file \tmtexttt{powheginput.f}. \ The
statement\\
\\
\tmtexttt{ \ \ \ \ \ rvalue=powheginput('myparm')}\\
\\
returns the value of token \tmtexttt{myparm} stored in
\tmtexttt{powheg.input}. If the token is not found in the input file, 
a message is printed, and the program is stopped.
The file is read only once, on the first invocation of the function
\tmtexttt{powheginput}, and token-value pairs are stored in internal arrays,
so that subsequent calls to \tmtexttt{powheginput} are relatively fast.
With the statement\\
\\
\tmtexttt{ \ \ \ \ \ rvalue=powheginput('\#myparm')}\\
\\
in case the token \tmtexttt{myparm} is not present, the program does not stop,
and returns the value~$- 10^6$.
The file \tmtexttt{powheginput.f} is a standalone code, and can be linked to
any program. In this way, an SMC that is reading an event file
may get parameters of the POWHEG-hvq run, if
it needs too.

We document here a typical input file \tmtexttt{powheginput.dat}:\\
\\
\tmtexttt{!Heavy flavour production parameters\\
\\
maxev 500000 ! number of events to be generated\\
ih1 1 ! hadron 1 type\\
ih2 -1 ! hadron 2 type\\
ndns1 191 ! pdf for hadron 1\\
ndns2 191 ! pdf for hadron 2}\\
\\
The integer \tmtexttt{ih1,ih2} and \tmtexttt{ndns1,ndns2} characterize the
hadron type and PDF used in POWHEG-hvq.
 The numbering scheme is
documented in the file \tmtexttt{hvqpdfpho.f}.
In that file, in the routine \tmtexttt{PRNTSF}, all PDF
sets available and the corresponding set number are listed. In particular, 191
corresponds to \tmtexttt{MRST2002}. The hadron type in \tmtexttt{ih1,ih2}
is 0 for a nucleon (i.e. the average of a proton and a neutron),
1~(-1)
for a proton~(antiproton), 2~(-2) for a
neutron~(antineutron), and 3~(-3) for a $\pi^+$~($\pi^-$).
Thus \tmtexttt{ih1=1,ih2=-1} corresponds to proton-antiproton
collisions.\\
\\
\tmtexttt{ebeam1 980 ! energy of beam 1\\
ebeam2 980 ! energy of beam 2}\\
\\
We assume that beam 1 and 2 move along the third axis in the positive
and negative direction respectively.\\
\\
\tmtexttt{qmass 175 ! mass of heavy quark in GeV\\
facscfact 1 ! factorization scale factor: mufact=muref*facscfact\\
renscfact 1 ! renormalization scale factor: muren=muref*renscfact\\
bbscalevar 1 ! if 0 use muref=qmass in Bbar calculation}\\
\\
Factorization and renormalization scale factors appearing here have to do with
the computation of the inclusive cross section
(i.e. the $\bar{B}$ function \cite{Nason:2004rx,Nason:2006hf,fnr2006}),
and can be varied by a factor of order 1 in both directions to study scale
dependence. Normally the reference scale is set equal to the transverse mass
of the heavy quark in the rest frame of the $q \bar{q}$ system.
If \tmtexttt{bbscalevar} is set to 0, the reference scale is chosen equal to
the heavy quark mass.
Other choices require a modification of the
the subroutine \tmtexttt{setscalesbb} in the file \tmtexttt{physpar-hvq.f}.
\\
The following is only needed if the quark is a top, and we want it to decay
with the inclusion of spin correlations\\
\\
\tmtexttt{topdecaymode 20000 ! an integer of 5 digits representing the decay
mode}.\\
\\
The value of the token is formed by five digits, each representing the maximum
number of the following particles in the (parton level) decay of the $t
\bar{t}$ pair: $e^{\pm}$, $\mu^{\pm}$, $\tau^{\pm}$, $u^{\pm}$, $c^{\pm}$.
Thus, for example, 20000 means the $t \rightarrow e^+ \nu_e b$, $\bar{t}
\rightarrow e^- \bar{\nu}_e \bar{b}$, 22222 means all decays, 10011 means one
of the top quarks
goes into electron or antielectron, and the other goes into any hadron, 00022
means fully hadronic, 00011 means fully hadronic with a single charm, 00012
fully hadronic with at least one charm. If all digits are 0, neither the $t$
nor the $\bar{t}$ are decayed.
Values that imply only one $t$ decay (for example 10000) are not implemented
consistently.
\\
In case \tmtexttt{topdecaymode} is different from 0, more parameters are needed
for the decay kinematics, and are used exclusively for decays\\
\\
\tmtexttt{tdec/wmass 80.4 \ \ \ ! W mass for top decay\\
tdec/wwidth 2.141 \ ! W width\\
tdec/bmass 5 \ \ \ \ \ \ ! b quark mass in t decay\\
tdec/twidth 1.31 \ \ ! top width\\
tdec/elbranching 0.108 \ ! W electronic branching fraction\\
tdec/emass 0.00051 ! electron mass\\
tdec/mumass 0.1057 ! mu mass\\
tdec/taumass 1.777 ! tau mass\\
tdec/dmass 0.100 \ \ ! d mass\\
tdec/umass 0.100 \ \ ! u mass\\
tdec/smass 0.200 \ \ ! s mass\\
tdec/cmass 1.5 \ \ \ \ ! charm mass\\
tdec/sin2cabibbo 0.051 \ ! sine of Cabibbo angle squared}\\
\\
Spin correlations in the decay are implemented, and effects due to the finite
width of the top and of the $W$ are also accounted for.
\\
The following parameters control the operation of the POWHEG-hvq program:\\
\\
\tmtexttt{! Parameters to allow-disallow use of stored data\\
use-old-grid 1\\
use-old-ubound 1}\\
\\
The meaning of these tokens requires a little knowledge of the operation of
POWHEG-hvq. Before the program starts generating events, the integral of the
inclusive cross section is computed, and a grid is set up for the generation of
Born-like configurations. Similarly, in the generation of hard 
radiation a grid is
computed to get an upper bounding function to the radiation probability. The
generation of the grids is time consuming, but the time spent in this way is
negligible in a normal run with hundreds of thousands of events being
generated. On the other hand, it is useful (for example, when debugging an
analysis program) to skip the generation stage. For this purpose, the grid for
the generation of Born-like kinematics is stored in the file
\tmtexttt{pwggrid.dat}.\\
If \tmtexttt{use-old-grid} is set equal to 0, 
and \tmtexttt{pwggrid.dat} exists and is
consistent, it is loaded, and the old grid and old value of the cross section
are used. Otherwise, a new 
grid is generated. Observe that the program does check
the file for consistency with the current run, but the check is not
exhaustive. The user should make sure that a
consistent grid is used.\\
The token \tmtexttt{use-old-ubound} has the same role as 
\tmtexttt{use-old-grid}, but it applies to the upper bounding
array that is used in the generation of radiation.
\\
The following parameters are used to control the grids generation\\
\\
\tmtexttt{! parameters that control the grid for Born variables generation\\
ncall1 10000 ! number of calls for initializing the integration grid\\
itmx1 5 ! number of iterations for initializing the integration grid\\
ncall2 100000 ! number of calls for computing integral and upper bound grids\\
itmx2 5 ! number of iterations for computing integral and upper bound grids\\
foldx 1 ! number of folds on x integration\\
foldy 1 ! number of folds on y integration\\
foldphi 1 ! number of folds on phi integration\\
! Parameters that controll the generation of radiation\\
nubound 100000 ! number of bbarra calls to setup upper bounds for radiation\\
iymax 1 !<=10, number of intervals in y grid to compute upper bounds\\
ixmax 1 !<=10, number of intervals in x grid\\
xupbound 2 ! increase upper bound for radiation generation by given factor}\\
\\
The values of some of the tokens may be changed in the following cases:
\begin{itemize}
  \item If the integration results have large errors, one may try to increase
  \tmtexttt{ncall1}, \tmtexttt{itmx1}, \tmtexttt{ncall2}, \tmtexttt{itmx2}.
  
  \item If the fraction of negative weights is large, one may increase
  \tmtexttt{foldx}, \tmtexttt{foldy}, \tmtexttt{foldphi}.
  Allowed values are 1, 2, 5, 10, 25, 50. The speed of the program is
  inversely proportional to the product of these numbers, so that a
  reasonable compromise should be found.

  \item If there are too many upper bound violation in the generation of
  radiation (see Section \ref{sec:cntstat}),
  one may increase \tmtexttt{nubound}, and/or \tmtexttt{xupbound}.
  
  \item If the efficiency in the generation of radiation is too small, one may
  try to increase \tmtexttt{iymax}, \tmtexttt{ixmax}.
\end{itemize}
In oder to check whether any of these conditions occurs,
the user should inspect
the file \tmtexttt{pwgstat.dat} at the end of the run, as illustrated in
sec.~\ref{sec:cntstat}.
\section{Examples}

Examples of \tmtexttt{powheg.input} files are given in the directories
\tmtexttt{c-tev}, \tmtexttt{b-tev}, \tmtexttt{t-tev}, \tmtexttt{tdec-tev} and
\tmtexttt{c-lhc}, \tmtexttt{b-lhc}, \tmtexttt{t-lhc}, \tmtexttt{tdec-lhc}.
In the examples in the \tmtexttt{tdec-tev,tdec-lhc} directories the $t \bar{t}$
pair is decayed semileptonically into an electron and an antielectron by
POWHEG-hvq.  In the \tmtexttt{t-tev,t-lhc}
directories the top decay is handled by
the SMC, according to its own parameters,
but without including spin correlations.
In all examples, the choice of the parameters that control the grid
generation is such
that a reasonably small fraction of negative weights is generated.
Even in the extreme case of charm production at the LHC, where
a fraction of negative weight less than 3\%{} is achieved.

\section{Counters and statistics}\label{sec:cntstat}

Several results relevant to the interpretation of the output of the run are
written to the file \tmtexttt{pwgstat.dat}. The fraction of negative weights,
the total cross section, the number of upper bound failures in the generation
of the inclusive cross section, and the generation efficiency, together with
failures and efficiency in the generation of hard radiation, are printed there.
These numbers are sufficient to take action in case of problems.

A call to the subroutine \tmtexttt{printstat} causes a printout of all
\tmtexttt{POWHEG-hvq} counters in \tmtexttt{pwgstat.dat} file.

\section{Using the PDF sets}\label{sec:pdf}

POWHEG-hvq uses the PDF implementation of \tmtexttt{hvqpdfpho.f}.
In that file, in the routine \tmtexttt{PRNTSF} all pdf
sets available, and the corresponding set numbers are listed.
In order to use a set, the corresponding
data file must be present in the directory where the run is performed.
If the file is
not found a message is printed with the name of the missing file, and the
program stops. In this case one should copy the missing file from the directory
\tmtexttt{pdfdata\tmtexttt{}} to the current directory,
or set up a symbolic link to it.

\section{Random number generator}
POWHEG-hvq uses the \tmtexttt{RM48} random number generator,
documented in the CERNLIB
writeups. This generator has default initialization. If a user wishes to start
the program with different seeds, he should call {\tt rm48in(iseed,n1,n2)}
(before calling the \tmtexttt{pwhginit} routine) in
order to seed the generator with the integer {\tt iseed}, and skip the first
{\tt n1+n2*10**8} numbers, as documented in the CERNLIB manual. This affects 
the
POWHEG-hvq random number sequence. If the program is interfaced to an SMC,
the user should also take care to initialize the seeds of the latter.


\begin{thebibliography}{10}
\bibitem{Frixione:2003ei}
  S.~Frixione, P.~Nason and B.~R.~Webber,
  ``Matching NLO QCD and parton showers in heavy flavour production,''
  JHEP {\bf 0308} (2003) 007
  [arXiv:hep-ph/0305252].

\bibitem{Nason:1988xz}
  P.~Nason, S.~Dawson and R.~K.~Ellis,
  ``The Total Cross-Section for the Production of Heavy Quarks in Hadronic
  Collisions,''
  Nucl.\ Phys.\  B {\bf 303} (1988) 607.
\bibitem{Nason:1989zy}
  P.~Nason, S.~Dawson and R.~K.~Ellis,
  ``The One Particle Inclusive Differential Cross-Section for Heavy Quark
  Production in Hadronic Collisions,''
  Nucl.\ Phys.\  B {\bf 327} (1989) 49
  [Erratum-ibid.\  B {\bf 335} (1990) 260].
\bibitem{Mangano:1992jk}
  M.~L.~Mangano, P.~Nason and G.~Ridolfi,
  ``Heavy quark correlations in hadron collisions at next-to-leading order,''
  Nucl.\ Phys.\  B {\bf 373} (1992) 295.
\bibitem{Nason:2004rx}
  P.~Nason,
  ``A new method for combining NLO QCD with shower Monte Carlo algorithms,''
  JHEP {\bf 0411} (2004) 040
  [arXiv:hep-ph/0409146].
\bibitem{Nason:2006hf}
  P.~Nason and G.~Ridolfi,
  ``A positive-weight next-to-leading-order Monte Carlo for Z pair
  hadroproduction,''
  JHEP {\bf 0608} (2006) 077
  [arXiv:hep-ph/0606275].


\bibitem{fnr2006}
  S. Frixione, P. Nason and G. Ridolfi, 
  ``A Positive-Weight Next-to-Leading-Order Monte Carlo for
  Heavy Flavour Hadroproduction'',
  arXiv:0707.3088 [hep-ph].
  
\bibitem{Frixione:2007zp}
  S.~Frixione, E.~Laenen, P.~Motylinski and B.~R.~Webber,
  ``Angular correlations of lepton pairs from vector boson and top quark decays
  in Monte Carlo simulations,''
  JHEP {\bf 0704} (2007) 081
  [arXiv:hep-ph/0702198].

\bibitem{Frixione:2006gn}
  S.~Frixione and B.~R.~Webber,
  ``The MC@NLO 3.3 event generator,''
  arXiv:hep-ph/0612272.

  \bibitem{Madevent}
  ``MadEvent: Automatic event generation with MadGraph''
  F. Maltoni and T. Stelzer, JHEP 0302:027, 2003
  [hep-ph/0208156] 
  
\bibitem{Boos:2001cv}
  E.~Boos {\it et al.},
  ``Generic user process interface for event generators,''
  arXiv:hep-ph/0109068.

\bibitem{Alwall:2006yp}
  J.~Alwall {\it et al.},
  ``A standard format for Les Houches event files,''
  Comput.\ Phys.\ Commun.\  {\bf 176} (2007) 300
  [arXiv:hep-ph/0609017].

\bibitem{Altarelli:1989wu} T. Sj\"ostrand et~al., in
  ``Z physics at LEP1: Event generators and software,'',  eds.
  G.~Altarelli, R.~Kleiss and C.~Verzegnassi, Vol 3, pg. 327.

\end{thebibliography}

\end{document}